# Freeing electrons from extrinsic and intrinsic disorder yields band-like transport in n-type organic semiconductors

M.-A. Stoeckel[1,§], Y. Olivier[2,§], M. Gobbi[1], D. Dudenko[2], V. Lemaur[2], M. Zbiri[3], A. Y. Guilbert[4], G. D'Avino[5], F. Liscio[6], N. Demitri[7], X. Jin[8], Y.-G. Jeong[8], M.-V. Nardi[9], L. Pasquali[10,11,12], L. Razzari[8], D. Beljonne[2*], P. Samorì[1*], E. Orgiu[1,8*]

[1] Université de Strasbourg, CNRS, ISIS, 8 allée Gaspard Monge, 67000 Strasbourg, France
[2] Laboratory for Chemistry of Novel Materials, University of Mons, Place du Parc, 20, B-7000 Mons, Belgium
[3] Institut Laue-Langevin, 71 Avenue des Martyrs, 38000 Grenoble, France
[4] Centre for Plastic Electronics and Department of Physics, Blackett Laboratory, Imperial College London, London SW7 2AZ, United Kingdom
[5] Institut Néel-CNRS and Université Grenoble Alpes, BP 166, F-38042 Grenoble Cedex 9, France
[6] CNR - IMM Sezione di Bologna, Via P. Gobetti 101, 40129 Bologna, Italy
[7] Elettra - Sincrotrone Trieste, S.S. 14 Km 163.5 in Area Science Park, I-34149 Basovizza, Trieste, Italy
[8] INRS-Centre Énergie Matériaux Télécommunications, 1650 Blv. Lionel-Boulet, J3X 1S2 Varennes, Québec
[9] Istituto dei Materiali per l'Elettronica ed il Magnetismo, IMEM-CNR, Sezione di Trento, Via alla Cascata 56/C, Povo, 38100 Trento, Italy
[10] Istituto Officina dei Materiali, IOM-CNR, s.s. 14, Km. 163.5 in AREA Science Park, 34149 Basovizza, Trieste, Italy
[11] Dipartimento di Ingegneria E. Ferrari, Università di Modena e Reggio Emilia, via Vivarelli 10, 41125 Modena, Italy
[12] Department of Physics, University of Johannesburg, PO Box 524, Auckland Park, 2006, South Africa

[§] These authors contributed equally to the work

*Corresponding authors: samori@unistra.fr; david.beljonne@umons.ac.be ; emanuele.orgiu@emt.inrs.ca

*Abstract*: **Charge transport in organic semiconductors is notoriously extremely sensitive to the presence of disorder, both intrinsic and extrinsic, especially for n-type materials. *Intrinsic* dynamic disorder stems from large thermal fluctuations both in intermolecular transfer integrals and (molecular) site energies in weakly interacting van der Waals solids and sources transient localization of the charge carriers. The molecular vibrations that drive transient localization typically operate at low-frequency (< a-few-hundred cm$^{-1}$), which renders it difficult to assess them experimentally. Hitherto, this has prevented the identification of clear molecular design rules to control and reduce dynamic disorder. In addition, the disorder can also be *extrinsic*, being controlled by the gate insulator dielectric properties. Here we report on a comprehensive study of charge transport in two closely related n-type molecular organic semiconductors using a combination of temperature-dependent inelastic neutron scattering and photoelectron spectroscopy corroborated by electrical measurements, theory and simulations. We provide unambiguous evidence that *ad hoc* molecular design enables to free the electron charge carriers from both intrinsic and extrinsic disorder to ultimately reach band-like electron transport.**

High charge carrier mobility is a prerequisite to ensure efficient electronic devices such as field-effect transistors (FETs), solar cells, and light-emitting diodes.[1,2] The charge carrier mobility in organic semiconductors results from the interplay of a complex set of physical parameters related to morphology, energetic and structural disorder, and defects, which ultimately are all intimately related the molecular chemical structure.[3] When resorting to organic molecular materials with high degree of crystallinity, the effects on charge transport of morphology and defects are minimized, providing tools to single out the role of disorder. A primary source of disorder is *intrinsic* and stems from large thermal fluctuations of (diagonal) site energies and, mostly, the (off-diagonal) intermolecular interactions mediating transport.[4] The other source of disorder for charge carriers in FETs is *extrinsic,* and it arises from the coupling of charge carriers to substrate phonons and randomly oriented dipoles at the semiconductor/gate dielectric interface.[5–7] Extrinsic disorder was invoked in previous reports as a key limiting factor to charge delocalization.[7] In spite of the above-mentioned intrinsic and extrinsic sources of disorder owing to the non-covalent and weak nature of the forces that hold together such van der Waals solids, several highly-performing molecular organic single crystals were found to exhibit *band-like transport* near room temperature. When properly supported by Hall measurements, the latter transport regime indicates that electron wave functions can be delocalized over several molecular units.[8–10] Although the sole standard electrical FET characterization is not capable of measuring the extent of charge delocalization, the band-like transport is usually invoked when the temperature dependence of the field-effect mobility ($\mu_{FET}$) resembles that observed in inorganic semiconductors such as silicon, i.e. showing an increase in $\mu_{FET}$ upon cooling down from room temperature.

Hence, one can confidently conclude that the combination of intrinsic and extrinsic disorder ultimately dictates the extent of charge delocalization in crystalline molecular semiconductors.

However, so far these two disorder components could never be disentangled. In addition, there is still considerable controversy as to which degree can a carrier wave function extend over neighboring molecules and especially how this is directly linked to the molecular structure. Here we use two different perylene diimide (PDI) derivatives as model systems to disentangle the role of the intrinsic vs. extrinsic disorder on electron transport in organic crystals. Through a systematic structural, electrical and spectroscopic characterization combining field-effect transistor devices, inelastic neutron scattering, and low-wavenumber spectroscopy measurements together with numerical simulations, we were able to isolate the most important vibrational lattice modes (for charge transport) of both PDI single crystals and to correlate them with the molecular chemical structure.

The two PDI derivatives (whose chemical structure is portrayed in Fig. 1a, 1b) are functionalized with cyano groups in the bay-region and feature similar environmental stability.[11–14] They only differ by the lateral chain on the imide position, initially designed to enhance their solubility. The first derivative, N-N'-bis(n-octyle)-(1,7&1,6)-dicyanoperylene-3,4:9,10-bis(dicarboximide) known as PDI8-$CN_2$, bears an alkyl chain on each side,[15] while its fluorinated derivative, N-N'-bis(perfluorobutyl)-(1,7&1,6)-dicyanoperylene-3,4:9,10-bis(dicarboximide) also called as PDIF-$CN_2$, exposes a fluorinated one[16,17].

Solution-processed single crystals were grown by means of solvent induced precipitation (SIP) (See SI), which consists in adding a small solution volume to a larger quantity of a non-solvent. Such a processing method enables to promote intermolecular interactions compared to the molecule-solvent ones, leading to the formation of highly pure crystals that precipitate at the bottom of the vial. The crystals were then drop-casted on an octadecyl trichlorosilane (OTS)-treated $SiO_2$ substrate (Fig. S4). The OTS functionalization acts as a molecular spacer that separates the dielectric surface from the PDI molecules sitting atop. More specifically, such dielectric treatment allows to decrease the coupling between the conjugated electron-transporting part of the molecules and the dielectric[7] and to prevent the presence of silanol groups at the interface of the surface that act as traps for electrons[18]. These traps are rendered inefficient when hydroxyl groups at the SiO$_2$ surface are passivated through the formation of a covalently bonded self-assembled monolayers. Two electrodes were evaporated through a shadow mask (Fig. S5) to form the final bottom-gate top-contact field-effect transistors (Fig. 1c, d, g). With this approach, we fabricated a top-contact bottom-gate solution-processed single-crystal FET without any contaminations resulting from the use of photoresist and its developer, which may occur in a standard lithographic process (Fig. S6). Furthermore, the dielectric functionalization (operated by functionalizing SiO$_2$ with OTS) allowing to disentangle extrinsic vs. intrinsic disorder effects on the electrical transport characteristics can only be employed in a bottom-gate architecture. As mentioned above, extrinsic disorder can be reduced by increasing the distance of the π-conjugated core of the molecular layers from the dielectric substrate. In previous studies, it has also been suggested that the polarizability tensor of individual rod-like conjugated molecules is anisotropic and it is larger along the longest molecular axis and smaller in the other spatial directions.[19] Because of their standing-up organization with their long molecular axis approximately perpendicular to the substrate, the rod-like conjugated core of both PDI derivatives studied here could be potentially subjected to stronger polarization effects from the dielectric. However, the lateral chains acting as a spacer make such contributions negligible for PDI derivatives while this has been attributed as the possible culprit for band-like transport not being observed in pentacene and sexithiophene single crystal FETs.[20] As confirmed by structural characterization, the conjugated core-dielectric distance is nearly identical in both PDIs, which also allows to consider nearly identical the amount of extrinsic disorder the two types of molecules are subjected to. Furthermore, PDI8-CN$_2$, and PDIF-CN$_2$ crystals were drop-casted on an OTS layer, which further ensures a minimization of the substrate-induced extrinsic disorder.

Both compounds present ideal n-type characteristics (Supplementary Table 1), with mobility values for PDIF-CN$_2$ being larger than for PDI8-CN$_2$ by about one order of magnitude over the temperature range investigated (Fig. 1 f, i). The trend of the temperature-dependent mobility (Fig. 1f) suggests band-like transport in PDIF-CN$_2$, as previsouly confirmed by Hall measurements.[7] This behavior could be generally observed in high-purity crystals and it is generally accompanied by an experimental trend of mobility, i.e. $\frac{\partial \mu}{\partial T} < 0$, when charge carrier mobility is not drain-bias dependent[21,22]. Below a given temperature, typically ranging between 180 and 220 K, the transport follows again a thermally-activated mechanism, although this latter point still needs a more accurate experimental validation which is beyond the scope of this work. Regarding PDI8-CN$_2$ (Fig. 1i), the transport seems to follow a purely thermally-activated behavior over the whole measured temperature range (80 K – 300 K). The charge transport in PDI8-CN$_2$ is characterized by an exponential dependence of the field-effect mobility with temperature.[23] Such behavior is characteristic of localized charge carriers that have to hop from one energy (molecular) site to a neighboring one in order to take part to the transport. Since the transport measurements on both PDI derivatives were carried out in three-terminal devices, the contact resistance as a function of temperature was measured (through a classical transmission line method) and the curves corrected (Fig. S7). The contact resistance was found to be larger in PDI8-CN$_2$ than in PDIF-CN$_2$ devices. This finding is fully consistent with the characterization performed by means of electron energy loss spectroscopy (EELS, Fig. 2a) and ultraviolet photoelectron spectroscopy (UPS, Fig. 2b, 2c) and in single crystals, being therefore fully comparable with the device case, which uses single crystals as the active layer. In particular, EELS measurements of the optical band gap (E$_{opt}$) of the crystals of both compounds provided nearly identical values for PDI8-CN$_2$ and PDIF-CN$_2$. While E$_{opt}$ and presumably the electronic band gap (HOMO-LUMO) of PDI8-CN$_2$ and PDIF-CN$_2$ are nearly equivalent, their respective ionization energies (Fig. 2b and Table S4) were found to be offset by as much as 0.70 eV (amounting to 7.9 eV and 7.4 eV, for PDIF-CN$_2$ and PDI8-CN$_2$, respectively). Considering the measured energy difference between E$_F$ and HOMO level (Fig. 2c), it appears clear that electron injection is certainly more favorable in PDIF-CN$_2$ thanks to the reduction of the energy barrier between Au work function and its LUMO. Structural analysis on the crystals confirmed the bulk structure, slipped-stacked for PDI8-CN$_2$ and brick-wall for PDIF-CN$_2$ (Fig. 2 d-i), with only one molecule per primitive cell, which is uncommon for small molecule semiconducting materials.[1,8,24] To rule out any possible phase transition upon temperature, we performed temperature-dependent structural characterization (see Section 4, Supplementary Information) and temperature-dependent solid-state F-NMR (Supplementary Fig. S13 and S14) on the fluorine atoms. No significant structural changes (phase transitions) were observed with both techniques with varying the temperature of the measurement. Thus, we are confident that the observed charge transport behaviors are not driven by any structural change.

In order to analyze intrinsic *static* disorder generated by the presence of trap states in the crystals, the trap density was extracted in both compounds in the framework of the space-charge-limited current analysis (Fig. S8).[25,26] PDI8-CN$_2$ crystals were found to exhibit a trap density (3 x 10$^{12}$ cm$^{-3}$) comparable to that of PDIF-CN$_2$ crystals (1 x 10$^{13}$ cm$^{-3}$). This further experimental evidence allows to state more soundly that the significant differences in transport regimes observed in our molecular systems do not stem from a difference in trap density but should be sought after in the intrinsic *dynamic* disorder behavior.

As announced earlier in the text, one way to reduce dynamic disorder consists in reducing the impact of molecular vibrations on the electronic structure, which corresponds to hampering the effect of both local and non-local electron–phonon couplings.[27] This can be accomplished by reducing the amplitude of the intermolecular displacements, measured up to 0.5 Å at T$_{amb}$.[28] Coupling of electrons to phonons yields modulation of the corresponding transfer integrals (by about the same order of magnitude as the transfer integral itself) and site energies (by up to 0.2 eV), thereby destroying the translational symmetry and collapsing the carrier wavefunctions.[29] Provided that the relevant time scale for charge motion is substantially shorter than the intermolecular vibrational time, charge carriers do transiently (de)localize in space. Hence, the frequency of the intermolecular vibrations is a crucial parameter. In particular, low-frequency (< 200 cm$^{-1}$) large-amplitude vibrations or vibrations displaying strong local and non-local electron-phonon coupling constants are those mostly hampering efficient charge transport.[30]

In order to investigate the charge localization degree of both PDI derivatives, we have performed atomistic calculations of the time-dependent electronic structure of the two molecular crystals by means of a combined classical/quantum modelling approach.[31] Computational details are provided in Supporting Information. In a nutshell, we first ran molecular dynamics (MD) simulations to sample the thermal lattice motion and to compute, as a function of time, the microscopic parameters governing the electron transport in both PDI derivatives. This includes intermolecular charge transfer integrals and molecular site energies, the latter explicitly accounting for supramolecular electrostatic interactions that largely contribute to energetic disorder.[32] We then fed this atomistic information into a tight-binding model for electron states in the two-dimensional high-mobility planes, in order to obtain the localization length (thermally-averaged inverse participation ratio) of electron carriers as a function of time.

The results of the hybrid classical/quantum calculations shown on Figure 3a point to a transient (de)localization of the electron wave function that breathes around average values of ~11 molecules in PDIF-CN$_2$ compared to only ~4 molecules in PDI8-CN$_2$. The large (small) averaged spatial extent of the electron carrier in PDIF-CN$_2$ (PDI8-CN$_2$) confirms the measured band-like (hopping-like) temperature-dependent mobility. This is a direct consequence of the crystalline packing of both derivatives, being slipped-stack for PDI8-CN$_2$ but brick-wall for PDIF-CN$_2$, as evidenced by the transfer integral distributions in Figure S16. In particular, compared to PDI8-CN$_2$, PDIF-CN$_2$ shows larger average transfer integrals, more balanced values along the pi-pi and pi-edge directions, and ultimately a smaller relative energetic disorder, i.e. the ratio between the fluctuations of transfer integrals and their mean value.

In order to gain a deeper understanding on the different charge transport mechanisms observed on PDI derivatives, we measured the low-frequency vibrations of both compounds. We employed *temperature-resolved* inelastic neutron scattering (INS) in combination with modeling and simulations. INS makes it possible to explore quantitatively phonon dynamics over the whole Brillouin zone, without being subjected to any specific constraint of selection rules. Unlike very recent works[33,34] based on low-temperature Stokes INS spectra measurements providing insight into molecular vibrations (internal modes), in the present study we used cold-neutron time-of-flight spectroscopy (see dedicated INS section in the SI), in order to measure in the up-scattering regime the anti-Stokes phonon modes (external modes) with a high-energy resolution and an excellent signal-to-noise ratio. Therefore, the low-frequency vibrations (up to 600 cm$^{-1}$) covering both crystal lattice modes (phonons or external modes) and some of the subsequent low-frequency molecular vibrations (internal modes) can be mapped out properly, allowing to study their temperature-dependence (150 - 300K) with direct implications on charge transport (Figure 3 b and 3 c). In the case of PDI8-CN$_2$, the vibrational peaks become more resolved upon lowering the temperature. This is expected due to a reduction of the thermally-induced displacements, related to the Debye-Waller factor, leading to less broadened features in inelastic neutron scattering. Interestingly, for PDIF-CN$_2$, upon cooling, some low-energy modes below 250 cm$^{-1}$ become clearly distinguishable, exhibiting a pronounced temperature-dependence, seemingly triggering/inducing other effects beyond (or in addition to) the expected reduction of thermal displacements. Further, a new vibrational feature appears at 545 cm$^{-1}$ in the generalized density of states g$^n$(E).[35] It must be emphasized here that the INS spectra are dominated by the dynamical degrees of freedom of hydrogen atoms, being the strongest neutron scatterers in PDIF-CN$_2$ and PDI8-CN$_2$. Hydrogens are located on the core of the materials in PDIF-CN$_2$ and both on the core and side chains in PDI8-CN$_2$. Therefore, INS intensities of the two PDIs are, in principle, not directly comparable. To underpin our INS measurements, we calculated the INS spectra by Fourier transform of the velocity autocorrelation functions, dissecting the contribution from individual elements along the MD trajectory. The partial contributions had to be neutron-weighted.[35] MD[36] simulations are found to be in a good agreement with the experimental INS spectra by applying a frequency scaling factor of 0.84 and 0.75 for PDI8-CN$_2$ and PDIF-CN$_2$, respectively (Figure 3d and 3e). Furthermore, the partial contributions show that the peak around 200 cm$^{-1}$ stems from the nitrogen contribution (Figure S24), with its intensity being larger in the case of PDI8-CN$_2$, and this is also reflected in the room-temperature THz spectroscopy spectrum (Figure S25). Carbon atoms form the host lattice to which the other atoms are bond. Therefore, their dynamical contribution reflects the main lattice vibration in terms of frequency spread. The calculated partial density of states of carbons highlight a stronger intensity of phonons for PDI8-CN$_2$ as compared to PDIF-CN$_2$, up to 192 cm$^{-1}$ (Figure S24). By combining INS with MD simulations, as shown in Figure 3d and 3e, it was possible on the one hand to validate the force field, and on the other hand to get a deeper insight into the partial lattice dynamical contributions. This enabled us to ultimately compare the INS intensities of the two PDIs.

As our model was able to reproduce quite confidently the whole vibrational modes measured by INS, we next proceeded with the calculation of the electron-phonon coupling spectral density for the transfer integrals along the pi-pi direction (see Figure 4a), as the temperature dependence is expected to be mainly sourced by non-local electron-phonon coupling along the dominant transport pathway. Results obtained for the off-diagonal coupling along the pi-edge (see Figure S17) direction as well as for the site energies (Figure S18) are reported in SI. We note that the intense peak around 400 cm$^{-1}$ for PDI8-CN$_2$ is also observed in the electron-phonon coupling spectrum of the site energies (see Figure S18).

These computed spectra convincingly show that low-energy (< 200 cm$^{-1}$) phonons are much more strongly coupled to the electronic degrees of freedom along the pi-pi direction in PDI8-CN$_2$ compared to PDIF-CN$_2$ (Figure 4b), thereby rationalizing the more pronounced trend towards breaking of the translational symmetry and spatial confinement of the electrons in the former molecule. The more pronounced intensity of the phonon spectra for PDI8-CN2 was experimentally proved also by room-temperature THz spectroscopy (Figure S25), which allows an immediate and quantitative comparison between the two PDI derivatives. In PDI8-CN$_2$, such low-energy phonons modulate the wave function overlap as a result of a combination of mostly short-axis translations and rotations[37,38] changing the distance and orientation along the packing direction (see SI for detailed assignment), thereby largely affecting the magnitude of the transfer integrals (in contrast with longitudinal displacements). The large (small) relative thermal molecular motions predicted for PDI8-CN$_2$ (PDIF-CN$_2$), as reported in Table S3, translate into broad (narrow) distributions in transfer integrals (Fig. S16). Altogether, similarly to didodecyl-benzothienobenzothiophene derivatives[39] and in contrast to the slipped-stack crystal organization of PDI8-CN$_2$, the brick-wall crystal organization in PDIF-CN$_2$ conveys both large electronic interactions and a smaller sensitivity to thermal energetic disorder that act in concert to delocalize the charge carriers, hence triggering band-like transport.

In summary, our systematic and comparative investigation of electron transport in single crystals based on PDI derivatives with subtle difference in the side chains revealed the existence of markedly different charge transport mechanisms, i.e. band-like vs. thermally activated transport. The careful design of our experimental work allowed to experimentally disentangle the effect of extrinsic vs. intrinsic disorder, and to focus on the relationship between the latter and the observed band-like transport through a number of (temperature-resolved) techniques. In particular, charge transport in PDIF-CN$_2$ and PDI8-CN$_2$ was analyzed through temperature-dependent electrical and structural characterization, corroborated by THz spectroscopy and by temperature-dependent inelastic neutron scattering. Our findings are also backed by a sound and fine atomistic modelling, which revealed a more pronounced degree of wave function delocalization for PDIF-CN$_2$ as a result of lower electron-phonon coupling. By focusing on the role of intrinsic disorder, our work also tackles one of the grand challenges in organic electronics, i.e. design rules for molecular semiconductors with as high as possible degree of electron delocalization. The experimentally observed differences in transport mechanism between band-like for PDIF-CN$_2$ and thermally-activated transport for PDI8-CN$_2$ can be ascribed to the different phononic activity at low wavenumber, as evidenced by electron-phonon coupling calculations. These calculations show that intrinsic dynamic disorder builds up in a qualitatively different way in brickwall and slipped-stacked PDI derivatives with respect to herringbone lattices of elongated molecules (e.g. pentacene, BTBT, etc). While in the latter case long-axis intermolecular displacements (below 50 cm$^{-1}$) provide a largely dominant contribution the total energetic disorder,[34] we have revealed here that the scenario is more complex for PDI, with several modes (up to 200 cm$^{-1}$, including short-axis translation and rotations) coming into play. As such, our work suggests rational molecular design guidelines for high mobility (n-type) molecular semiconductors: a combined effect of translations and rotations along the short molecular axis gives rise to low-wavenumber phonons capable of modulating the electron wave function overlap and to affect the magnitude of the transfer integrals in a much more pronounced manner than molecular longitudinal displacements.

More specifically, we found that large relative thermal molecular motions, as those predicted for PDI8-CN$_2$, should be minimized when designing new molecular compounds as they translate into broad distributions in transfer integrals. In our study, we show an example of how engineering the side chains and their relative interactions (-CF$_3$ vs. -CH$_3$ groups), leads to different packing motifs that allow electron wave functions to be delocalized over more than 10 molecular units. Hence, our experimental results coupled with simulations can be used to screen electron and hole conducting molecular semiconductors to predict a set of high-performance materials (exhibiting band-like transport), therefore significantly contributing to a more refined chemical design space for next-generation electronics materials. Although this work puts forward the importance of controlling vibrational dynamics through molecular design for electron charge transport in organic FETs, studying and engineering phonons in organic semiconductors is also key for exciton separation in OPVs and the control of heat transport in organic-based thermoelectrics.

## Methods

*X-Ray Diffraction traces as a function of temperature*

Single crystal X-ray diffraction measurements were performed on several PDIF-CN$_2$ crystals at XRD1-ELETTRA beamline. Diffraction images were collected at different temperatures: 100 K, 150 K, 175 K, 200 K, 250 K and 300 K. The crystal structure was confirmed to be triclinic *P*-1 with one molecule per cell as reported in literature.[16] During thermal treatment, lattice parameters (*a*, *b*, *c*, *α*, *β*, *γ*) slightly changed. The strongest variation was the expansion by 3% of the *c*-axis at 300 K, which lead the volume to increase by 4.4%. These results do not suggest a real phase transition, however some minor molecular rearrangement inside the unit cell can be supposed due to the thermal expansion. The powder diffraction of PDI8-CN$_2$ were collected during the thermal annealing. Some Bragg peaks splitting and shifts were observed in the diffraction patterns during thermal treatment, indicating a small anisotropic contraction of the unit cell. The variation was found to be smaller than that observed in PDIF-CN$_2$ crystals.

CCDC 1848556, 1848559, 1848555, 1848558, 1848557 and 1848554 contain the supplementary crystallographic data for compounds PDIF-CN$_2$ at 100 K, 150 K, 175 K, 200 K, 250 K and 300 K. These data can be obtained free of charge from The Cambridge Crystallographic Data Centre via https://www.ccdc.cam.ac.uk/structures.

*Ultraviolet Photoelectron Spectroscopy – Electron Energy Loss Spectroscopy*

The as received samples were mounted on a sample holder and inserted in the analysis chamber (base pressure 1 x 10$^{-10}$ mBar) for the analysis. Spectra were collected with a double pass cylindrical mirror analyzer (PHI-15-255GAR) driven at constant pass energy. Energy resolution was set to 0.1 eV. UPS was taken using a He discharge lamp from VG (photon energy 21.2 eV). The EELS spectra were collected using an electron beam at 160 eV and at a current of 10 nA.

*MAS-ssNMRas a function of temperature*

Fluorine Solid state NMR experiments were done on an AVANCE 500 MHz wide bore spectrometer (BrukerTM) operating at a frequency of 470.45 MHz for 19F and equipped with a triple resonance MAS probe designed for 2.5 mm o.d. zirconia rotors (closed with Kel-F caps) and a BCU extreme for temperature regulation. In order to properly simulate spectra undistorted, lineshapes are needed, therefore all spectra were acquired with the original Hahn's echo sequence[40] (without 1H decoupling). This echo was synchronized with the MAS rotation and set equal to two rotation periods for the different MAS speeds (ranging from 15 to 26 kHz). FIDs were sampled over 8192 time-domain points spaced by 1 μs Dwell time, leading to a 122.07 Hz and 500 kHz (1062.8142 ppm) spectral resolution and width respectively. 16 scans were added for full phase cycle completion and noise averaging. A Lorentzian line broadening of 150 Hz was applied prior to Fourier transformation that was done on 16384 points (zero fill to 16 K). The referencing was done by setting 19F PTFE signal at -122 ppm (room temp).

The spectra were treated with Topspin software. Through the deconvolution process of each peak, the chemical shift, the chemical shift anisotropy (CSA) and the integral can be extracted as relevant information. The spectrum generated by the deconvolution process is compared to the measured one with an indication of the percentage of overlapping between them. For all the analysis, the lowest value of overlapping reached is 93.86%, for an average of 96.1%.

From that fitting, 6 chemically equivalent atoms are found, due to the position on the chain and the interaction with the atoms of the neighbor molecules. In case of phase transition, since the environment of the atoms should change, the CSA should change as well or additional peaks would appear. However, no particular changes can be observed at low temperature. The fluctuations observed after 220 K are coming from the reduction of the spin rate to 15.140 kHz, which increase the incertitude of the deconvolution.

*Temperature-dependent Inelastic Neutron Scattering*

The temperature-dependent inelastic neutron scattering (INS) measurements were performed using the direct geometry cold neutron, time-focusing time-of-flight spectrometer IN6 at the Institut Laue-Langevin (ILL) (Grenoble, France). About 250 mg of powder samples of PDIF-$CN_2$ and PDI8-$CN_2$, prepared as described above, were sealed inside thin flat Aluminum holder that was fixed to the sample stick of a cryofurnace. Data were collected at 150, 200, 300 and 420 for PDIF-$CN_2$, and at 200, 300 and 420 for PDI8-$CN_2$. The INS spectra were collected in the up-scattering regime (neutron energy-gain mode) using the high-resolution mode and a neutron incident wavelength $\lambda_i$=4.14 Å ($E_i$ = 4.77 meV), corresponding to a maximum Q ~ 2.6

Å$^{-1}$ on IN6, and offering a good resolution within the considered dynamical range for the anti-Stokes data. Standard corrections including detector efficiency calibration and background subtraction were performed. The data analysis was done using ILL procedures and software tools. The *Q*-averaged, one-phonon, generalized phonon density of states (GDOS) was obtained using the incoherent approximation[41–43]. In this context, the measured scattering function *S*(*Q*, *E*), as observed in the INS experiments, is related to the phonon generalized density of states *g$^n$(E)*, as seen by neutrons, as follows:

$$g^{(n)}(E) = A < \frac{e^{2W_i(Q)}}{Q^2} \frac{E}{n_T(E) + \frac{1}{2} \pm \frac{1}{2}} S(Q,E) > \qquad (1)$$

With:

$$g^{(n)}(E) = B \sum_i \left\{\frac{4\pi b_i^2}{m_i}\right\} x_i g_i(E) \qquad (2)$$

where the + or – signs correspond to energy loss or gain of the neutrons respectively and n$_T$(E) is the Bose-Einstein distribution. *A* and *B* are normalization constants and $b_i$, $m_i$, $x_i$, and $g_i(E)$ are, respectively, the neutron scattering length, mass, atomic fraction, and partial density of states of the i$^{th}$ atom in the unit cell. The quantity between < > represents suitable average over all *Q* values, within the high-Q ranges indicated above, at a given energy. 2*W*(*Q*) is the Debye-Waller factor. The weighting factors $\frac{4\pi b_i^2}{m_i}$ for various atoms in the units of barns/amu are[44] : H: 81.37, C: 0.46, N: 0.82, O: 0.26, and F: 0.21.

*Modelling and simulation*

*Molecular Dynamics simulations:* Molecular Dynamics (MD) simulations were performed using a re-parameterized version of the Dreiding force field.**[45]** After a NVT equilibration dynamics, we have run a NVT production dynamics of 5 ps at 298 K, saving configurations of the system every 5 fs. The total simulation time and the time bin for sampling were chosen to allow for accurate sampling of low-frequency phonons modes. MD simulations have been performed with the Materials Studio (MS) 6.0 code.[46]

*Microelectrostatic calculations:* We have computed the contribution from intermolecular electrostatic interactions to the site energies, including charge-multipole and charge-induced dipole interactions, by means of a classical atomistic self-consistent microelectrostatic model implemented in MESCAL[47] code.

*Transfer integrals calculations:* The transfer integrals between nearest-neighbour PDI dimers along the pi-pi and pi-edge directions have been calculated using the fragment approach, as implemented in the Amsterdam Density Functional (ADF) package.[48] Because of the non-orthogonality of the fragment (or monomer) basis set, a Löwdin transformation to the initial electronic Hamiltonian is applied resulting in the following expression of the transfer integral $\tilde{t}_{12}$:

$$\tilde{t}_{12} = \frac{t_{12} - (\varepsilon_1 + \varepsilon_2)S_{12}}{1 - S_{12}}$$

Where $t_{12}$ and $\varepsilon_i$ are the transfer integral and the site energies calculated in the fragment basis set and $S_{12}$ the fragment orbitals overlap.

*Site energies and transfer integrals electron-phonon coupling calculations:* The site diagonal and off-diagonal electron-phonon coupling spectra have been obtained by a Fourier transform of the site energy and transfer integrals autocorrelation functions using MATLAB 7.11.

*Localization length and tight binding model:* From the polarization energy and transfer integrals calculated along the MD trajectory, we have built a time-dependent tight-binding model considering one LUMO orbital per molecular site. Solving this Hamiltonian yields time-dependent electronic eigenstates from which we assess the degree of (de)localization through the computed Boltzmann-averaged inverse participation ratio (see definition in Supplementary materials section).

Supplementary material

CCDC 1848556, 1848559, 1848555, 1848558, 1848557 and 1848554 contain the supplementary crystallographic data for compounds PDIF-CN$_2$ at 100 K, 150 K, 175 K, 200 K, 250 K and 300 K. These data can be obtained free of charge from The Cambridge Crystallographic Data Centre via https://www.ccdc.cam.ac.uk/structures.

**Acknowledgments**

E. O. is supported by the Natural Sciences and Engineering Research Council of Canada (NSERC) through an individual Discovery Grant. This work was financially supported by EC through the ERC project SUPRAFUNCTION (GA-257305), the Marie Curie ITN projects BORGES (GA No. 813863) and UHMob (GA- 811284), the Labex projects CSC (ANR-10-LABX-0026 CSC) and NIE (ANR-11-LABX-0058 NIE) within the Investissement d'Avenir program ANR-10-IDEX-0002-02, and the International Center for Frontier Research in Chemistry (icFRC). The Institut Laue-Langevin (ILL) facility, Grenoble, France, is acknowledged for providing beam time on the IN6 spectrometer. The work in Mons was supported by the European Commission / Région Wallonne (FEDER – BIORGEL project), the Consortium des Équipements de Calcul Intensif (CÉCI), funded by the Fonds National de la Recherche Scientifique (F.R.S.-FNRS) under Grant No. 2.5020.11 as well as the Tier-1 supercomputer of the Fédération Wallonie-Bruxelles, infrastructure funded by the Walloon Region under Grant Agreement n1117545, and FRS-FNRS. The research in Mons is also funded through the European Union Horizon 2020 research and innovation program under Grant Agreement No. 646176 (EXTMOS project). DB is a FNRS Research Director.

**Author contributions**

E.O. conceived the research. E.O., P.S. and D.B. supervised the work. M. –A. S. fabricated the devices and performed the electrical measurements under the guidance of M. G. and E. O.. M. –A. S. carried out the solid-state NMR measurements. Y.O., D. O., V. L. and G. D. performed the simulations. A. Y. G. and M. Z. carried out the INS experiments, analyzed the data and assisted with data interpretation. F.L. and N. D. carried out the XRD experiments and analyzed the data. M. N. and L. P. performed the UPS and EELS measurements and analyzed the data. X.-J., Y. –G. J. and L. R. carried out the THz measurements and analyzed the data. E.O., M.-A. S. and P.S. wrote the manuscript and all authors participated in manuscript preparation and editing.

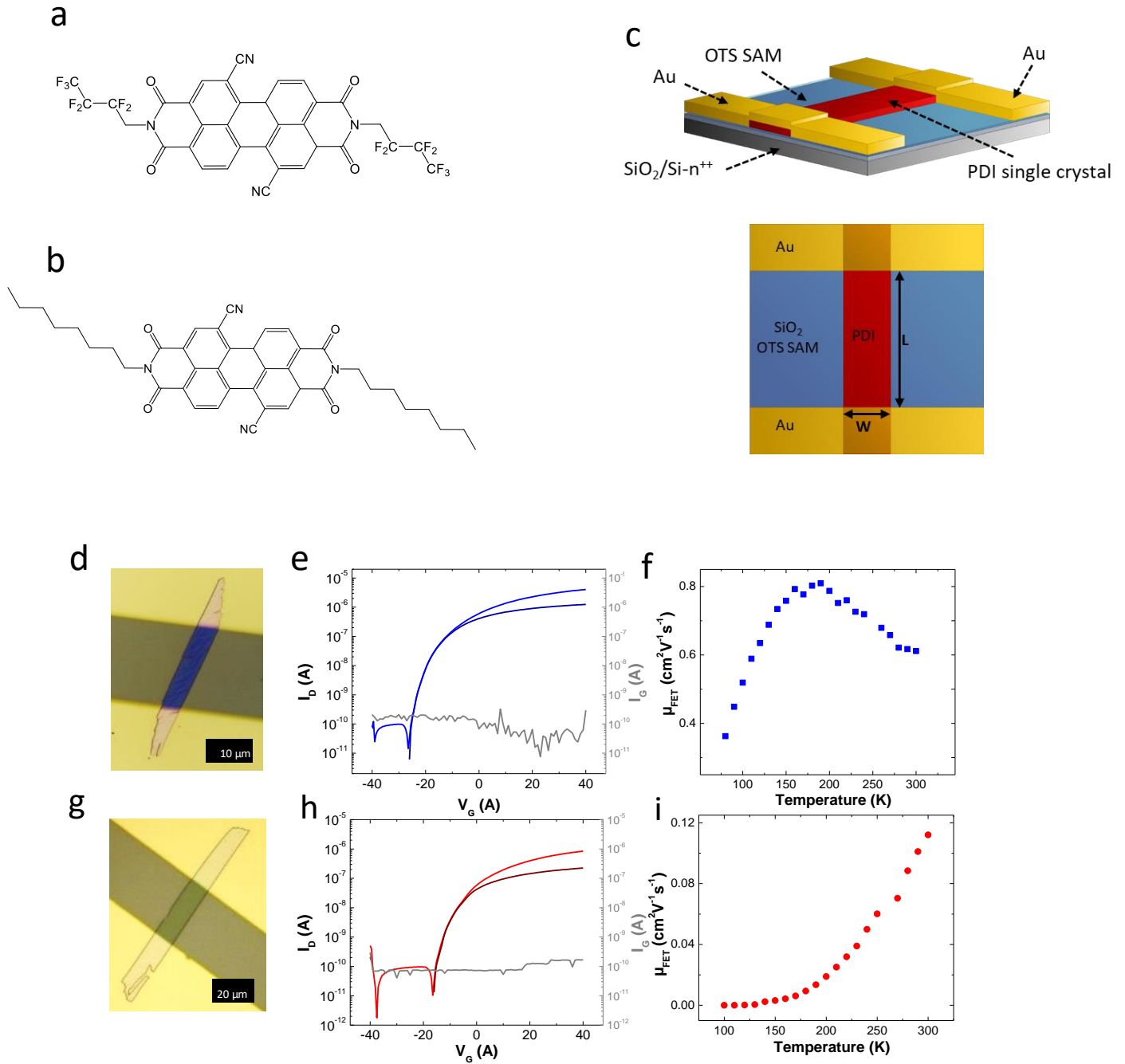

**Figure 1: Temperature-dependent electrical characterization of PDI-based devices.**
**a**, chemical structure of PDIF-CN$_2$. **b**, chemical structure of PDI8-CN$_2$ **c**, Schematics of a device based on PDI single crystal, 3D and top view. **d**, optical microscopy image of a PDIF-CN$_2$ single-crystal FET, **e**, transfer curve of a single-crystal FET of PDIF-CN$_2$ with $V_D = 10$ V (in dark blue) and $V_D = 40$ V (in blue). The gate current is plotted in grey. **f**, evolution of the mobility of a PDIF-CN$_2$ sample as a function of temperature, exhibiting band-like behavior. **g**, optical image of a PDI8-CN$_2$ single-crystal FET **h,** transfer curve of a PDI8-CN$_2$ single-crystal FET with $V_D = 10$ V (in dark red) and $V_D = 40$ V in red. **i**, evolution of the mobility of a PDI8-CN$_2$ sample as a function of temperature, revealing thermally-activated transport.

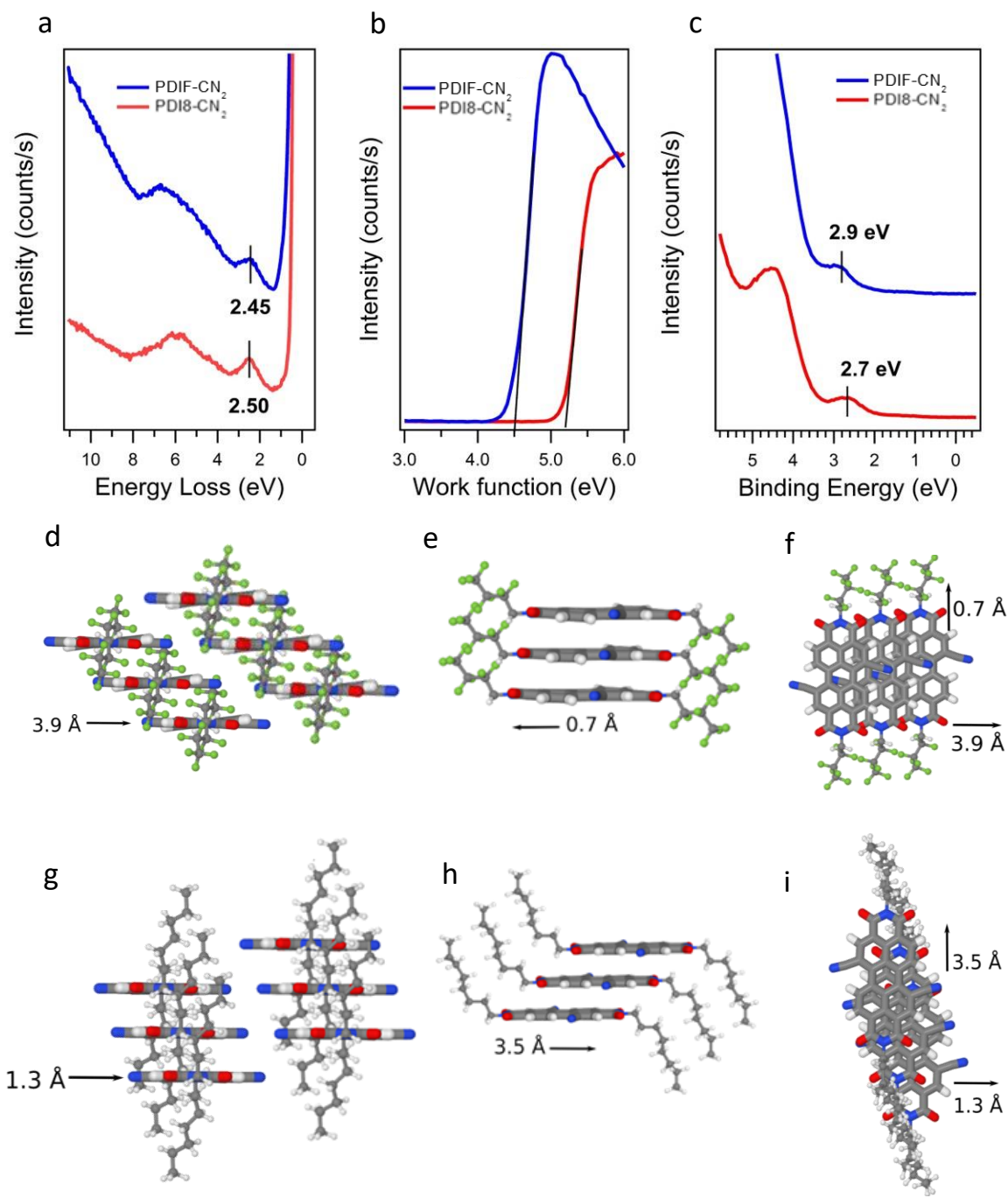

**Figure 2: Electronic and structural properties of PDIF-CN$_2$ and PDI8-CN$_2$**
**a**, EELS spectra of PDIF-CN$_2$ and PDI8-CN$_2$. The marked values in the experimental traces represent the energy difference between the primary beam and the energy of the first loss structure i.e. the optical band gap (E$_{opt}$). E$_{opt}$ corresponds to 2.50 eV for PDIF-CN$_2$ and 2.45 eV for PDI8-CN$_2$. **b** and **c**, work function and E$_F$-HOMO measurements for PDIF-CN2 and PDI-CN$_2$, respectively. **d, e, f** side view and top view packing of PDIF-CN$_2$. **g, h, i** side view and top view packing of PDI8-CN$_2$.

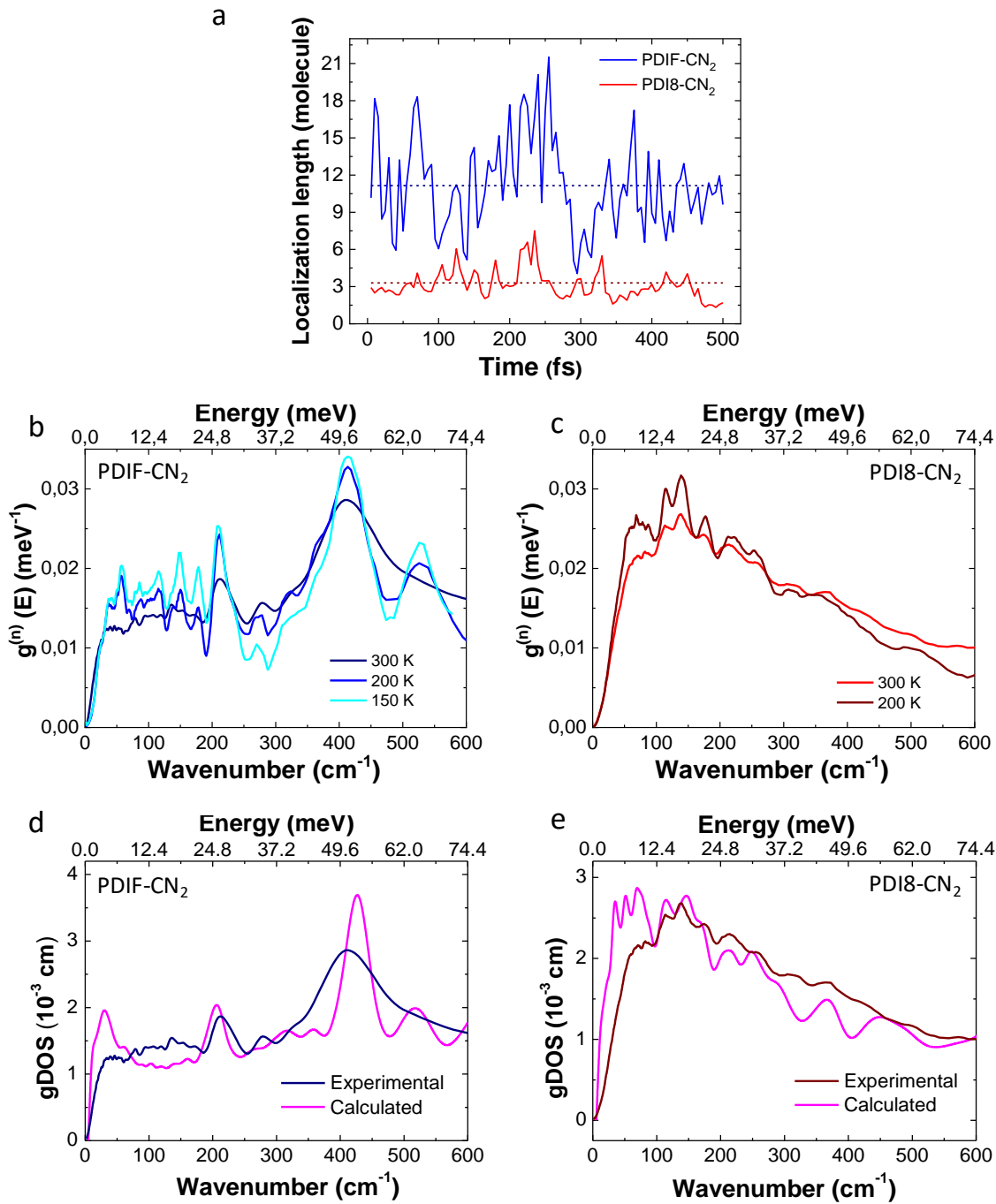

**Figure 3: Experimentally determined and calculated vibrational modes of PDIF-CN$_2$ and PDI8-CN$_2$**

**a**, Localization length over time: charge carriers are delocalized over ca. 11 PDIF-CN$_2$ molecules (in blue). Conversely, charge carriers are localized only over ca. 4 molecular units in the case of PDI8-CN$_2$ (in red). **b**, and **c**, temperature-dependent inelastic neutron scattering spectra of PDIF-CN$_2$ and PDI8-CN$_2$ respectively. (Full INS spectra resolved in temperature shown in S22); **d** and **e,** comparative experimental-simulated INS spectra of PDIF-CN$_2$ and PDI8-CN$_2$, respectively (room temperature).

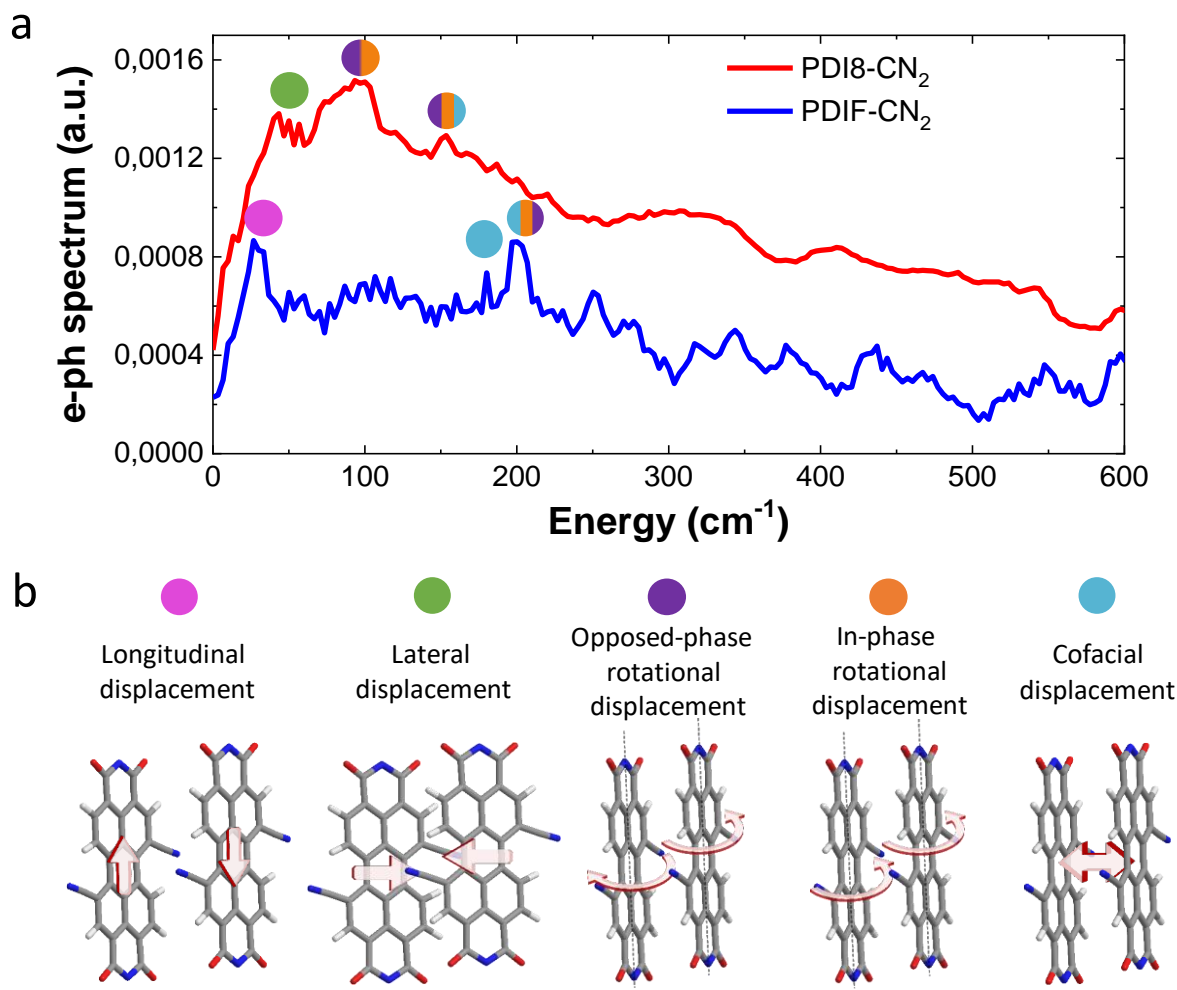

**Figure 4: Electron-phonon coupling analysis and assigned molecular displacement for PDI8-CN$_2$ and PDIF-CN$_2$**

**a,** Electron-phonon coupling spectrum of the time-dependent transfer integrals along the pi-pi direction in crystals of PDI8-CN$_2$ (in red) and PDIF-CN$_2$ (in blue), respectively (T = T$_{amb}$). **b,** Different molecular displacements and related color code associated to the spectra in (a).